\documentclass[12pt]{iopart}
\usepackage{iopams}
\usepackage{graphicx}
\usepackage[dvips]{color}
\usepackage{amssymb, amsfonts, amsbsy}
\usepackage{mathrsfs} 

\def\d{{\mathrm{d}}}
\def\implies{\Rightarrow}

\newcommand{\bt}{\bar{t}}
\newcommand{\oL}{\omega_{L}}
\newcommand{\oE}{\omega_{E}}

\newcommand{\oEdot}{\dot{\omega}_{E}}
\newcommand{\oLdot}{\dot{\omega}_{L}}

\def\lint{\hbox{\Large $\displaystyle\int$}} 
\def\eg{{\it e.g.}}

\def\eof{\Box}

\begin{document}
\def\TODAY{19 December 2008; 8 January 2009; 05 September 2009; 28 November 2009}
\title[Signature change events: A challenge for quantum gravity?]{Signature change events: \\
A challenge for quantum gravity?}
\author{Angela White$^1$, Silke Weinfurtner$^2$, and
Matt Visser$^3$}
\address{$^1$ School of Mathematics and Statistics, Newcastle University,\\
\ \ Newcastle upon Tyne, NE1 7RU, UK}
\address{$^2$  Department of Physics and Astronomy, University of British Columbia, \\
\ \ Vancouver, BC, V6T 1Z1, Canada}
\address{$^3$ School of Mathematics, Statistics, and Operations Research, \\
\ \ Victoria University of Wellington, PO Box 600, Wellington, New Zealand}
\ead{ang.c.white@gmail.com,
silke@physics.ubc.ca,
matt.visser@msor.vuw.ac.nz}
\begin{abstract}
Within the framework of either Euclidian (functional-integral) quantum gravity or canonical general relativity the signature of the manifold is \emph{a priori} unconstrained. Furthermore, recent developments in the emergent spacetime programme have led to a physically  feasible implementation of (analogue) signature change events. This suggests that it is time to revisit the sometimes controversial topic of signature change in general relativity. Specifically, we shall focus on the behaviour of a quantum field defined on a manifold containing regions of different signature. We emphasise that, regardless of the underlying classical theory, there are severe problems associated with any quantum field theory residing on a signature-changing background. (Such as the production of what is naively an infinite number of particles, with an infinite energy density.)  We show how the problem of
 quantum fields exposed to finite regions of Euclidean-signature (Riemannian) geometry has similarities with the quantum barrier penetration problem. Finally we raise the question as to whether signature change transitions could be fully understood and dynamically generated within (modified) classical general relativity, or whether they require the knowledge of a theory of quantum gravity.
\vskip 0.250cm
\noindent
Keywords: signature-change, Euclidean signature, no boundary proposal, Bogoliubov coefficients, cosmological particle production, barrier penetration.
\vskip 0.250cm
\noindent \TODAY;  \\ \LaTeX-ed \today.
\end{abstract}


\maketitle
\tableofcontents
\markboth{Signature change events: A challenge for quantum gravity?}{}
\clearpage
\markboth{Signature change events: A challenge for quantum gravity?}{}

\def\d{{\mathrm{d}}}
\section{Introduction}
Signature change in general relativity is a subject that over the years has sometimes attracted considerable heated controversy. Kinematically, the key point is that the choice of signature, the pattern of the \emph{signs} of the eigenvalues of the matrix of metric coefficients, is both a coordinate invariant, and is an independent choice that is made simply for observational physical reasons. The signature does not seem to be constrained at the level of dynamics~\cite{Ellis-et-al, Dray-et-al}. Neither the Einstein--Hilbert action nor the Einstein equations are tied to a specific choice of signature.   This feature of general relativity becomes most obvious in two places: in Euclidean quantum gravity (where the spacetime manifold is indeed formally continued to Euclidean signature)~\cite{EQG, no-boundary, Hartle-Hawking, arrow}, and in the canonical Hamiltonian formulation of (ordinary Einstein--Hilbert) classical general relativity (where the overall signature of the spacetime is determined by the behaviour of a \emph{non-dynamical} part of the metric related to the ``lapse function'')~\cite{Teitelboim}.  Indeed, it is always possible to locally pick coordinates to foliate any arbitrary manifold in the form
\begin{equation}
\d s^2 = - B \; \d t^2 + g_{ij} \, (\d x^i + N^i \, \d t) \, (\d x^j + N^j\, \d t).
\end{equation}
Here $N^j$ is the shift vector, and $B$ is simply some real function of position. If $B$ is positive, then $B=N^2$, where $N$ is the lapse function, and the manifold has Lorentzian signature. In contrast, if  $B$ is negative, then $B=-N^2$, where $N$ is now the ``Euclidean lapse'' function, and the manifold has Riemannian signature. (The coordinates themselves, $t$ and $x^i$, always remain real.)

In analogue spacetimes (emergent spacetimes) the quantity $B$ is proportional to the $(speed)^2$ of whatever signal one is interested in studying~\cite{ergoregion, LRR}. Specifically, in BECs this is the $(speed)^2$ of the phonons, which is proportional to the so-called  ``scattering length'', which in turn can be externally controlled \emph{and driven negative} by using an external magnetic field to tune the system through a Feshbach resonance~\cite{LRR, Calzetta, silke+angela}. Thus analogue spacetimes can provide explicit physical models for implementing (some of the key features of) signature change in laboratory systems --- a circumstance that strongly indicates that it may be worthwhile to re-assess the whole notion of signature change in general relativity itself.

Specifically, how does a quantum field theory (QFT) residing on a spacetime undergoing signature change react to the imposition of a signature change event? A specific and relatively simple example of this phenomenon has been carefully studied by Dray \emph{et al.}~\cite{Dray-et-al}, who considered massless conformally coupled scalar fields in a two-dimensional spacetime undergoing a simultaneous universe-wide signature change process of the form 
\begin{equation}
(1+1) \to (0+2) \to (1+1).
\end{equation}
Here the $(a+b)$ notation denotes $a$ time dimensions and $b$ spatial dimensions in a region of constant signature.
Their key result can be slightly rephrased as follows: In this particular process, the Bogoliubov coefficients governing particle production are given (up to unimportant phases)  by the \emph{exact} expressions:
\begin{equation}
\alpha = \cosh\left( c_0 \,k\, \Delta \tau\right); \qquad \beta =  i\sinh\left( c_0\, k\, \Delta \tau\right);
\end{equation}
where $\Delta \tau$ is the total (Euclidean) proper time ($\Delta\tau = \int N \; \d t$) for which the universe exhibits Euclidean signature ($B<0$), $k$ is the wavenumber of the mode under consideration, $c_0$ is the physical speed of light, and the conformal flatness of arbitrary $(1+1)$ spacetimes was an essential ingredient in their original calculation. Additional background material can be found in references~\cite{Dray-other, Kandrup, controversy1, controversy2}. An immediate result of this calculation is that both the total number and total energy of the particles generated in such  $(1+1) \to (0+2) \to (1+1)$ signature change events is infinite~\footnote{We shall use $N_\#$ whenever we are talking about total particle number to avoid any possibility of confusion with the lapse function $N$.}:
\begin{equation}
N_\# = \oint |\beta|^2 \;\d k = \oint \left|\sinh\left( c_0\, k\, \Delta \tau\right)\right|^2 \; \d k = \infty;
\end{equation}
\begin{equation}
E = \oint c_0\, k\,  |\beta|^2 \;\d k = \oint c_0\, k\,  \left|\sinh\left( c_0\, k\, \Delta \tau\right)\right|^2 \; \d k = \infty.
\end{equation}
This feature of the calculation is disturbing, and one possible interpretation of these results is that placing a QFT on a manifold actually \emph{prevents} signature change --- this is similar to the situation regarding topology change where for some time it was similarly felt that infinite particle production by topology change events might suppress them completely~\cite{topology}.  (Current understanding is more subtle, and progress regarding topology change has been made by enlarging the ``universe of discourse'' to contain more general classes of spacetime~\cite{topology2}.)
Below we shall generalise and modify the Dray~\emph{et al.}~result in several ways:
\begin{itemize}

\item First, it is useful to verify that the basic result is not
significantly modified by going to $(d+1)$ dimensions, and in
particular the physically interesting $(3+1)$ dimensions.  Though
generally one no longer has conformal flatness, this condition can
for calculational purposes be replaced by a condition that  the
universe is not expanding during the  signature change event. (That is, the signature change event takes place on a hypersurface of zero extrinsic curvature.)

\item Second, we shall soon see that the inclusion of particle
masses, which is important for any realistic attempt at
model building based on this cosmological particle production
effect, also leads to a closed-form \emph{exact} result
\begin{eqnarray}
\alpha &=& \cosh\left\{ \sqrt{m^2 c_0^4 + c_0^2\,k^2} \;\Delta \tau\right\}; 
\\
\beta &=&  i\sinh\left\{ \sqrt{m^2 c_0^4 + c_0^2\,k^2} \; \Delta \tau\right\};
\end{eqnarray}
where again the universe is taken to be non-expanding during the signature change event.

\item Third, in situations where the universe is undergoing
expansion (or contraction) during the signature change event,
while one cannot generically write down \emph{exact} expressions
for the Bogoliubov coefficients, there are nevertheless
\emph{reliable} WKB  \emph{estimates} for these coefficients.
Namely, up to irrelevant phases:
\begin{eqnarray}
\alpha &\approx& \cosh\left\{ \lint_{\!\!\!\!\!\!E} \sqrt{m^2 c_0^4 + {c_0^2 \,k^2\over a(\tau)^2}} \; \;\d \tau  \right\};
\\
\beta &\approx&  i\sinh\left\{  \lint_{\!\!\!\!\!\!E} \sqrt{m^2 c_0^4 + {c_0^2 \,k^2\over a(\tau)^2}} \; \;\d \tau \right\};
\end{eqnarray}
where $a(\tau)$ is the scale factor of the universe as a function
of Euclidean proper time, and the integral runs only over the Euclidean
region $E$.

\item Fourth, we shall also consider the (nowadays
rather popular) possibility of an ultra-high-energy  breakdown of
Lorentz invariance~\cite{Mattingly, causal-sets, broken-BEC}, to
see what happens to the particle production spectrum in this situation. If the dispersion relation in the absence of cosmological expansion is some arbitrary function $\omega= \Omega(k)$, then in the presence of expansion the dispersion relation becomes $\omega = \Omega(k/a(\tau))$, and we shall (up to irrelevant phases) derive the estimates:
\begin{eqnarray}
\alpha &\approx& \cosh\left\{ \int_E  \; \Omega\!\left({k/ a(\tau)}\right) \;\d \tau  \right\};
\\
\beta &\approx&  i\sinh\left\{  \int_E  \; \Omega\!\left({k/ a(\tau)}\right) \;\d \tau \right\}.
\end{eqnarray}

\end{itemize}
In all of these situations we still see infinite particle production, $N_\#=\infty$, and infinite energy being dumped into the QFT modes, $E=\infty$. Thus the infinities encountered in the Dray~\emph{et al.}~calculation are not specific to their particular toy model, but are instead generic features of Lorentzian$\to$Euclidean$\to$Lorentzian signature change events, with a behaviour quite different from ordinary cosmological particle production~\cite{Parker}. This is a circumstance that certainly calls out for some careful and delicate analysis. 

For instance, a more drastic choice based on an analogy with BEC physics that additionally breaks time reparameterisation invariance, can be used to successfully regularise total particle production, though total energy release is still formally infinite~\cite{silke+angela}.
We emphasise that in the current article we are working strictly
within the confines of standard general relativity, and are not
adopting any ``analogue model'' point of view such as considered
in our earlier article~\cite{silke+angela}. (For general background on
these analogue model approaches see~\cite{ergoregion, LRR, analogue}.)  
This approach changes the entire focus and intent of the investigation, 
potentially making it more interesting to the bulk of the
relativity and cosmology communities. Staying within the confines of general relativity means we are \emph{not} working with a bi-metric theory and technically simplifies calculations as we have the full coordinate invariance of standard general relativity available to us.  Indeed, strategic use of the full coordinate invariance
of standard general relativity (and in particular the use of full time reparameterization invariance) is essential to keeping the calculations tractable.

In the discussion below we shall first look at the kinematics of signature
changing spacetimes, carefully specifying the precise nature of
the problem we wish to solve (in particular the relevant PDE and
junction conditions). We then consider certain exact results we can
derive for non-expanding but signature changing spacetimes,
follow this up with a discussion of signature change in an
expanding universe, and then turn to the question of a possible
ultra-high-energy breaking of Lorentz invariance. Finally in the conclusions we shall turn to the question of what this all tells us about the possibility of signature change in the observable universe --- is the apparent pathology of this curved-space QFT calculation telling us that we need to step outside the framework of semi-classical gravity? Is this calculation telling us something about quantum gravity itself?

\section{Kinematics of signature change in $(d+1)$ dimensions}

\subsection{Coordinate independent properties of signature change events\label{Sec:Coordinate.Independent.Sig.Changes}}

The matrix of metric coefficients $g_{ab}$ is a real symmetric $(d+1)\times(d+1)$ matrix which has $(d+1)$ eigenvalues. These eigenvalues are of course coordinate dependent, but the pattern of \emph{signs} of the eigenvalues is independent of one's choice of coordinates --- this is a classic result most typically referred to as Sylvester's theorem (or Sylvester's law of inertia)~\cite{Sylvester}. The pattern of signs of the eigenvalues of the matrix of metric coefficients is referred to as the \emph{signature} of the manifold. If one sign is different from all the others, this is taken to qualitatively indicate the ``time'' direction and the manifold is said to have Lorentzian signature. If all signs are the same the manifold is said to have Euclidean signature. (Lorentzian and Euclidean signature are the two most important cases.) If some eigenvalues are zero the manifold is said to have degenerate signature. (We shall take our manifolds to be almost everywhere non-degenerate. Degeneracies, if present, will be confined to isolated hyper-surfaces of signature change. Note that due to the coordinate invariance of general relativity, a coordinate reparameterization can be made to obtain a continuous signature change event.) The remaining cases are said to have Kleinian signature~\cite{Kleinian}, corresponding for instance to more than one time and more than one space directions. It is important to note that signature can (in the first instance) be defined in a purely kinematical way before even introducing any notion of dynamics --- of course if we then want to generalise the discussion to asking questions about a possible dynamical origin or explanation for signature change events, then consideration of the dynamics is unavoidable.

\subsection{Signature change events in FLRW spacetimes\label{Sec:FLRW}}

We are particularly interested in spatially flat ($k=0$) FLRW-like  spacetimes, where space and time are already decomposed in an appropriate manner, and thus the shift vectors are not necessary, $N^j = 0$. Standard FLRW geometries exhibit one degree of freedom, the scale factor; we shall write the scale-factor-squared as $A(t)$, and in principle allow $A(t)$ to change sign. In addition, we allow the lapse-squared to be an arbitrary function $B(t)$, also capable of  flipping sign
\begin{equation} \label{Eq:Canonical.FLRW}
\d s^2 = - B(t) \,\d t^2 + A(t) \,\d\vec x^2.
\end{equation}
For $B(t)=1>0$ and $A(t)=a(t)^2>0$ we recover the standard representation for a spatially flat ($k=0$) FLRW universe. More generally,  $B(t)$ and $A(t)$ are in principle allowed to go through zero and to change sign.
The relevant metric, inverse metric, and metric determinant are
\begin{equation}
\label{E:def1}
g_{ab} = \mathrm{diag}(-B,A,A,\dots);
\qquad
g^{ab} = \mathrm{diag}\left(-{1\over B},{1\over A},{1\over A},\dots\right);
\end{equation}
\begin{equation}
\label{E:def2}
g = - B A^d;
\qquad\qquad\qquad\qquad
\sqrt{-g} = \sqrt{ B A^d}.
\end{equation}
The coordinate speed of light, which we shall denote by $c$, (as opposed to the physical speed of light, which we shall denote by  $c_0$), is defined by
\begin{equation}
c^2 = B/A.
\end{equation}
At a Lorentzian$\to$Euclidean or Euclidean$\to$Lorentzian signature change event:
\begin{itemize}
\item $A(t)$ can in principle  change sign (as a function of $t$), but is fixed (in the sense that it is independent of how one coordinatises $t$).
\item $B(t)$ can change sign (as a function of $t$), and is variable under reparameterisation of $t\to\bar t = f(t)$:
\begin{equation}
B(t) \;\d t^2 = \bar B(\bar t) \; \d \bar t^2 \qquad\implies\qquad  \bar B = B \; \left({\d  t\over\d \bar t}\right)^2.
\end{equation}
\item
Note however that at any particular point in spacetime on the manifold, the \emph{sign} of $B$ is fixed (independent of how one coordinatises $t$).  That is, for any arbitrary foliation, the sign of $B$ is fixed for all points in the same hypersurface.
\end{itemize}
There are in principle (at least) 3 distinct types of signature change:
\begin{equation}
-++\dots \to +++\dots
\end{equation}
\begin{equation}
 -++\dots \to ---\dots
 \end{equation}
 \begin{equation}
 -++\dots \to +--\dots
\end{equation}
depending on whether $B$, $A$, or both change sign\;\footnote{We shall not, in the current article, consider  more exotic forms of signature change, such as Kleinien signature~\cite{Kleinian}, where one deals with two or generally more timelike directions (and the relevant physical questions become ones of barrier penetration and reflection).}. Only the first two cases are truly instances of Lorentzian$\to$Euclidean signature change, the third case corresponds to a sign flip in an overall conformal factor but does not change that nature of the light cones (there is still one ``special'' direction which we can call time, and $d$ ``other'' directions we can call space). The second case involves a Big Bang singularity as $A(t)\to0$, (and so $a(t)\to 0$), which while interesting in its own right is not germane to the discussion at hand.
So we shall only be interested in the case where $B(t)$ alone changes sign, that is when
\begin{equation}
-++\dots \to +++\dots \to -++\dots
\end{equation}
This will be our fundamental model of a Lorentzian$\to$Euclidean$\to$Lorentzian signature-change event.
Because of the coordinate invariance of general relativity one can parametrise  the $t$ coordinate $t\to\bar t$ so that quite generically and without any loss of generality:
\begin{itemize}
\item  We can choose the coordinate $\bar t$ so that $\bar B$ satisfies:
\begin{equation}
A^d/\bar B = 1/\epsilon = \pm 1.
\end{equation}
That is,
\begin{equation}
\epsilon = \mathrm{sign}[B(t)] = \mathrm{sign}[\bar B(\bar t)] .
\end{equation}
Equivalently,
\begin{equation}
 \bar B = \epsilon A^d  = B \; \left({\d t\over\d \bar t}\right)^2.
 \end{equation}
 \item
 This particular choice is made to make the d'Alembertian operator look simple in these coordinates. Specifically
\begin{equation}
\Delta_{d+1} = {1\over\sqrt{-g}} \partial_a \left( \sqrt{-g} \; g^{ab} \; \partial_b \; \_\_\_ \right).
\end{equation}
This becomes
\begin{equation}
\Delta_{d+1} = {\epsilon\over A^{d/2}} \left( {\partial\over \partial \bar t} \right)^2 + {\nabla^2\over A}.
\end{equation}
\item We also have
\begin{equation}
\left({\d  t\over\d \bar t}\right) = \sqrt{\epsilon A^d /B} = \sqrt{A^d/|B|}.
\end{equation}
Equivalently,
\begin{equation} 
\left({\d  \bar t\over\d  t}\right) = \sqrt{|B|} A^{-d/2}.
\end{equation}
\item For the coordinate speed of light we then have
\begin{equation}
c^2=B/A \qquad\implies\qquad \bar c^2 = \bar B/A = \epsilon \; A^{d-1}.
\end{equation}
\end{itemize}
This completes our specification of the \emph{kinematics} of the spacetime geometry we are interested in.

\subsection{Differential equation and junction conditions}

Note that we can always use the coordinate freedom of general relativity to set $\bar B = \pm A^d$ and thus apparently make the signature change ``discontinuous''.  This is however ``merely'' a choice of coordinates. A careful analysis (see reference~\cite{Ellis-et-al}, and related discussions in Dray~\emph{et~al.}~\cite{Dray-et-al}) agree on the following procedure:  To see what the junction conditions are it is sufficient to work at the classical level and consider for instance a massive  minimally coupled classical scalar field governed by the Klein--Gordon equation
\begin{equation}
{1\over\sqrt{-g}} \partial_a \left( \sqrt{-g} \; g^{ab} \; \partial_b \phi \right) = m^2 \phi.
\end{equation}
Now rewrite the Klein--Gordon equation in the explicit form
\begin{equation}
\partial_a \left(
\mathrm{diag}\left[ - \sqrt{A^d/B}, \sqrt{B A^{d-2}},  \sqrt{B A^{d-2}}, \dots \right]^{ab}\; \partial_b \phi
\right) = \sqrt{ B A^d} \; m^2 \; \phi,
\end{equation}
that is
\begin{equation}
- \partial_t \left( \sqrt{A^d/B} \; \partial_t \phi\right) + \nabla \cdot \left( \sqrt{B A^{d-2}} \; \nabla\phi \right) =  \sqrt{ B A^d} \; m^2 \; \phi.
\end{equation}
Fourier transforming in \emph{space} (but not time) this becomes
\begin{equation}
\partial_t \left( \sqrt{A^d/B} \; \partial_t \phi\right) = - \left[ \sqrt{ B A^d} \; m^2 +  \sqrt{B A^{d-2}} \; k^2 \right] \phi.
\end{equation}
Now use the coordinate invariance to transform $t\to\bar t$ and substitute back into the ODE, then
\begin{equation}
\partial_{\bar t} \left( \sqrt{1/\epsilon} \; \partial_{\bar t} \phi\right) = - \left[ \sqrt{ \epsilon A^{2d}} \; m^2 +  \sqrt{\epsilon A^{2d-2}} \; k^2 \right] \phi,
\end{equation}
that is, (certainly everywhere apart from those ``apparently discontinuous'' points where $\epsilon$ changes sign),
\begin{equation}
\partial_{\bar t} \left( \partial_{\bar t} \phi\right) = - \left[ \sqrt{ \epsilon^2 A^{2d}}\; m^2 +  \sqrt{\epsilon^2 A^{2d-2}} \; k^2 \right] \phi.
\end{equation}
We can rewrite this as
\begin{equation}
\label{E:KG-reduced}
\partial_{\bar t}^2 \phi = - \epsilon \; A^d \left[ m^2 +  k^2/A \right] \phi,
\end{equation}
where at worst we will have to apply some junction conditions at
the points where $\epsilon$ changes sign. This should be compared
with the Dray~\emph{et~al.}~calculation~\cite{Dray-et-al}, which
corresponds to the special case $m=0$, $d=1$.
Physically we can reinterpret this (still classical) calculation as a specific parametric oscillator
\begin{equation}
\partial_{\bar t}^2 \phi = - \bar \omega^2 \; \phi,
\end{equation}
with
\begin{equation}
\bar \omega^2 = \epsilon \; A^d \left[ m^2 +  k^2/A \right].
\end{equation}
There is a discontinuous jump in the ``potential'' at the point where $\epsilon$ changes sign:
\begin{itemize}
\item The Lorentzian regions are ``classically allowed''.
 \item
The Euclidean region is ``classically forbidden''. 
\item If you
think of $\bar \omega^2$ as a ``potential'',
then this potential has infinitely sharp walls where $\epsilon$
changes sign, and  $\bar \omega^2\to \pm |\bar
\omega|^2$. 
\item If one reinterprets this as a
quantum mechanical barrier penetration problem, then in the
Euclidean region one is attempting to tunnel through a barrier
with exactly \emph{one half} of the energy one would classically
need to just skim over the top of the ``jump''.
\end{itemize}
At the barrier walls
you will have to apply the Dray~\emph{et~al.}~boundary
conditions~\cite{Dray-et-al}:
\begin{equation}
\label{E:J-field}
 [\phi]=0;
 \qquad\qquad
 [ n^a \; \partial_a \phi] = 0.
\end{equation}
This second condition can be written in several equivalent forms. For instance,
\begin{equation}
\left[ {\partial\phi\over \partial \tau} \right] = 0,
\end{equation}
or, using
\begin{equation}
\d \tau = |c| \; \d t = \sqrt{|B|/A} \; \d t,
\end{equation}
we also have
\begin{equation}
\left[ \sqrt{A\over |B|} \; {\partial\phi\over \partial t} \right] = 0.
\end{equation}
But since we are assuming $A$ is continuous this is equivalent to asserting
\begin{equation}
\left[ \sqrt{A^d\over |B|} \; {\partial\phi\over \partial t} \right] = 0,
\end{equation}
or finally
\begin{equation}
\label{E:J-momentum}
\left[ {\partial\phi\over \partial \bar t} \right] = 0.
\end{equation}
This last form of the junction condition is the one that is actually easiest to use. (And this is the form of the junction condition that one might have most easily derived by simply integrating the differential equation across the signature change event.) 
Summarizing,  the differential equation we need to solve is
\begin{equation}
\partial_{\bar t}^2 \phi = - \epsilon \; |\bar\omega|^2 \;\phi,
\end{equation}
subject to the boundary conditions
\begin{equation}
\label{E:J-final}
 [\phi]=0;
 \qquad\qquad
 [\partial_{\bar t} \phi] = 0.
\end{equation}
Though for ease of presentation we have so far phrased things directly in terms of the PDE which governs the field oscillations, and so have implicitly phrased the analysis in a first-quantized wavefunction formalism, the same results can be obtained in a number of different ways (\emph{e.g.} we could rephrase everything in terms of continuity of the classical field and classical momentum density, or we could perform a second-quantized mode analysis). These junction conditions also carry through for curved-spacetime QFTs.

\section{Non-expanding universe\label{Sec:PPflat}}

Ultimately the differential equation we need to solve is
\begin{equation}
\partial_{\bar t}^2 \phi = - \epsilon \; |\bar\omega|^2 \;\phi,
\end{equation}
where in the current situation $\bar\omega$ is piecewise constant.
Let $\epsilon$ change sign at $\bar t_-$ and $\bar t_+$. Then in
the three regions of interest we have
\begin{eqnarray}
\bar t \leq \bar t_-: \qquad &  \psi& = \exp(-i|\bar\omega|\bar
t);
\\
\bar t \in (\bar t_-,\bar t_+): \qquad &  \psi& = A
\exp(+|\bar\omega|\bar t) + B  \exp(-|\bar\omega|\bar t);
\\
\bar t \geq \bar t_-: \qquad &  \psi& = \alpha
\exp(+i|\bar\omega|\bar t) + \beta \exp(-i|\bar\omega|\bar t).
\end{eqnarray}
Applying the junction conditions at the first signature changing
event we have
\begin{eqnarray}
\exp(-i|\bar\omega|\bar t_-)  &=& +A \exp(+|\bar\omega|\bar t_-) +
B  \exp(-|\bar\omega|\bar t_-);
\\
\exp(-i|\bar\omega|\bar t_-)  &=& i A \exp(+|\bar\omega|\bar t_-)
- i B  \exp(-|\bar\omega|\bar t_-);
\end{eqnarray}
and from again applying the junction conditions at the second
signature changing event we obtain
\begin{eqnarray}
\fl A \exp(+|\bar\omega|\bar t_+) + B  \exp(-|\bar\omega|\bar t_+)
&=& \alpha \exp(+i|\bar\omega|\bar t_+) + \beta
\exp(-i|\bar\omega|\bar t_+);
\\
\fl A \exp(+|\bar\omega|\bar t_+) - B  \exp(-|\bar\omega|\bar t_+)
&=& i\alpha \exp(+i|\bar\omega|\bar t_+) - i\beta
\exp(-i|\bar\omega|\bar t_+).
\end{eqnarray}
These are easily seen to yield
\begin{eqnarray}
A &=& {1+i\over2}  \exp(-|\bar\omega|\bar t_-)
\exp(+i|\bar\omega|\bar t_-);
\\
B&=& {1-i\over2} \exp(+|\bar\omega|\bar t_-)
\exp(+i|\bar\omega|\bar t_-) ;
\end{eqnarray}
leading to
\begin{eqnarray}
\alpha &=& \cosh\{ \,|\bar\omega| \;\, (\bar t_+-\bar t_-)\, ]\,\}
\; \exp\{\,-i|\bar\omega|\,  (t_+-t_-)\, \} ;
\\
\beta &= &  i \sinh\{ \,|\bar\omega| \;\, (\bar t_+-\bar t_-)\,
]\,\} \; \exp\{\,i|\bar\omega|\,  (t_+ + t_-)\, \}.
\end{eqnarray}
Up to irrelevant phases we can cast this as
\begin{eqnarray}
\alpha &\sim& \cosh\{ \,|\bar\omega| \; \Delta \bar t\,\};
\\
\beta &\sim &  i \sinh\{ \,|\bar\omega| \; \Delta \bar t\,\}.
\end{eqnarray}
But from the definition of $|\bar\omega|$, and inserting suitable
factors of $c_0$,
\begin{equation}
|\bar \omega| =  A^{d/2} \sqrt{ m^2 c_0^4+
c_0^2 \, k^2/A }\, .
\end{equation}
Remember that in the current calculation $A$ is a constant, $A^{d/2} \Delta \bar t = \Delta\tau$, and we furthermore 
can without loss of generality set $A\to1$ by rescaling the
spatial coordinates. Thus
\begin{eqnarray}
|\alpha| &=& \cosh\left\{ \,\sqrt{ m^2 c_0^4+  c_0^2 \, k^2} \; \Delta \tau\,\right\};
\\
|\beta| &= &  \sinh\left\{ \,\sqrt{ m^2 c_0^4+  c_0^2 \, k^2}\; \Delta \tau\,\right\}.
\end{eqnarray}
From a second-quantized viewpoint, this implies particle production given by
\begin{equation}
N_k = \sinh^2\left\{ \,\sqrt{ m^2 c_0^4+  c_0^2 \, k^2}\; \Delta \tau\,\right\}.
\end{equation}
These formulae generalize the results of Dray \emph{et al.}~\cite{Dray-et-al}, in that they are now applicable in any number of spatial dimensions, and also allow for particle mass.

%
\section{WKB estimate}
%
If $A(t)$ is not a constant, corresponding to an expanding (or
contracting)  spatially flat FLRW universe that undergoes a
signature change event, then for ``slowly expanding'' universes one can use WKB theory to give a
reliable estimate for the classical\,/\,quantum field excitations.
For a minimally coupled scalar in ($d+1$) dimensions undergoing a
$L\to E\to L$ signature change transition we found, in equation (\ref{E:KG-reduced})
that the Klein-Gordon equation reduces to:
\begin{equation}
\ddot \phi = - \epsilon A^d \left[ m^2 +  k^2/A \right] \phi.
\end{equation}
Here a choice of $\epsilon=+1$ corresponds to Lorentzian regions
of spacetime, and in the intermediate Euclidean region,
$\epsilon=-1$.  To estimate the particle production using WKB
theory, we seek an approximate solution to the Klein--Gordon equation in each region, and connect the solutions at each hypersurface of signature change
by applying the Dray~\emph{et al.}~junction conditions.
That is, the scalar field $\phi$ and its conjugate momentum $\partial_{\bar t} \phi$ are continuous
at a hypersurface of signature change --- see (\ref{E:J-field}) and (\ref{E:J-momentum}).

\subsection{Approximate solutions in the three regions}

We choose the initial Lorentzian region to extend from
$(-\infty,a)$.  The Euclidean region exists from the first
hypersurface of signature change at $\bt=a$, extending until the
second hypersurface of signature change $\bt=b$. The final
Lorentzian region occurs from $(b,+\infty)$.
Note this problem is mathematically identical to that of flux hitting a
steep-walled potential well. In this case the regions of
Lorentzian signature would correspond to the two classically
allowed regions, $(-\infty,a)$ and $(b,+\infty)$ while the
classically forbidden region corresponds to the Euclidean region.
The signature changing hypersurfaces play the same role as
(sharp wall) ``turning points'' if we are looking at a barrier penetration
problem.

$\bullet$ In the initial Lorentzian region, we start by defining $\omega_L = \sqrt{\bar \omega^2}$, and consider the approximate solution
\begin{equation}
\phi(\bt \leq a) = {\exp(-i\int_{\bt}^{a} \oL \d \bt)\over
\sqrt{\oL}}\,,
\end{equation}
with conjugate momentum
\begin{equation}
\dot{\phi}(\bt \leq a) = +i\sqrt{\oL}\;\exp\left(-i\int_{\bt}^a
\oL \d \bt \right) -{1\over2}{\oLdot\over \oL}\; \phi.
\end{equation}
Throughout the article, we use the notation $\dot{\gamma}=\partial_{\bt}
\gamma = \partial \gamma/\partial \bar t$.

$\bullet$ In the final Lorentzian region,  we again proceed by defining $\omega_L = \sqrt{\bar \omega^2}$, and the scalar field takes the (approximate) form
\begin{equation}
\phi(\bt\geq b) = \alpha   \;{\exp(+i\int_b^{\bt} \oL \d \bt)\over
\sqrt{\oL}} + \beta \; {\exp(-i\int_b^{\bt} \oL \d \bt)\over
\sqrt{\oL}}\, ,
\end{equation}
with conjugate momentum
\begin{equation}
\fl
\dot{\phi}(\bt\geq b) = i \alpha \; \sqrt{\oL}
\exp\left(+i\int_b^{\bt} \oL \d {\bt}\right) - i  \beta \; \sqrt{\oL}
\exp\left(-i\int_b^{\bt} \oL \d {\bt}\right) -
{1\over2}{\oLdot\over \oL} \; \phi \,.
\end{equation}

$\bullet$ In the Euclidean region, which extends from the first signature change hypersurface at $\bt=a$, up to the second signature change hypersurface at $\bt=b$,  we start by defining $\omega_E = \sqrt{-\bar \omega^2}$. Approximate solutions for the scalar field can be written as
\begin{equation}
\phi =  A \; {\exp\left(+\int_a^{\bt} \oE \d
{\bt}\right)\over\sqrt\oE} +  B \; {\exp\left(-\int_a^{\bt} \oE \d
{\bt}\right)\over\sqrt\oE}\,,
\end{equation}
with the ``conjugate momentum''  in the Euclidean region being given by
\begin{equation}
\fl \dot{\phi} =  A \; \sqrt\oE \;{\exp\left(+\int_a^{\bt} \oE \d
\bt\right)} -  B \; \sqrt\oE \;{\exp\left(-\int_a^{\bt} \oE \d
\bt\right)} -{1\over2} {\oEdot\over\oE}\; \phi \, .
\end{equation}

\subsection{Junction conditions at the two allowed-to-forbidden transitions:}
We now apply the Dray~\emph{et al.}~junction conditions at the two signature change events.
Note that at the two junctions we have
\begin{equation}
\omega_L(a^-) = \omega_E(a^+); \qquad \dot \omega_L(a^-) =  \dot \omega_E(a^+);
\end{equation}
\begin{equation}
\omega_E(b^-) = \omega_L(b^+); \qquad \dot \omega_E(b^-) =  \dot \omega_L(a^+).
\end{equation}
Before the first signature changing event, the scalar field takes
the form
\begin{equation}
\phi({\bt}\leq a) = {\exp(-i\int_{\bt}^a \oL \d {\bt})\over
\sqrt{\oL}} \, ,
\end{equation}
while just after the hypersurface of signature change, but still within
the Euclidean region,
\begin{equation}
\phi(a\leq {\bt}\leq b) =  A \; {\exp\left(+\int_a^{\bt} \oE \,\d
{\bt}\right)\over\sqrt\oE} +  B \; {\exp\left(-\int_a^{\bt} \oE \,\d
{\bt}\right)\over\sqrt\oE}\, .
\end{equation}
We connect the scalar field across the signature change event by
employing the junction conditions we expounded on previously,
these are, that both the scalar field and conjugate momentum are
continuous: $[\phi]=0$ and $ [\dot{\phi}] = 0$.
Continuity of the scalar field gives us
\begin{equation}\label{con1a}
1=A+B,
\end{equation}
while from the requirement that the conjugate momentum is continuous at
$a$, we obtain
\begin{equation}\label{con2a}
i = A - B.
\end{equation}
Therefore
\begin{equation}
\label{ABspecial}
A = {1+i\over2}; \qquad B = {1-i\over 2}\, .
\end{equation}
Just after the \emph{first}
signature changing event we have
\begin{equation}
\phi({\bt}\gtrsim a) =  {1+i\over2} \times {\exp\left(+\int_a^{\bt} \oE\, \d
{\bt}\right)\over\sqrt\oE} +  {1-i\over 2} \times {\exp\left(-\int_a^{\bt} \oE\, \d
{\bt}\right)\over\sqrt\oE},
\end{equation}
whence, near the \emph{second} signature changing event
\begin{eqnarray}
 \phi({\bt}\lesssim b) &=&  {1+i\over2} \times {\exp\left(+\int_a^b \oE \,\d
{\bt}\right) \exp\left(-\int_{\bt}^b \oE \,\d
{\bt}\right)\over\sqrt\oE} 
\nonumber\\
&&
+  {1-i\over 2}\times {\exp\left(-\int_a^b \oE \, \d
{\bt}\right) \exp\left(\int_{\bt}^b \oE \, \d
{\bt}\right)\over\sqrt\oE}\,.
\end{eqnarray}
Define $\Theta = +\int_a^b \oE \; \d {\bt}$,
so that just before the second signature changing event we have
\begin{equation}
\phi({\bt}\lesssim b) =  {1+i\over2} \; e^\Theta \;\;{ \exp\left(-\int_{\bt}^b
\oE \d {\bt}\right)\over\sqrt\oE} +   {1-i\over 2} \; e^{-\Theta} \;\;
{\exp\left(\int_{\bt}^b \oE \d {\bt}\right)\over\sqrt\oE}.
\end{equation}
We now need to match this to the scalar field in the final
Lorentzian region, that is
\begin{equation}
\phi({\bt}\geq b) = \alpha \;  {\exp(+i\int_b^{\bt} \oL \d
{\bt})\over \sqrt{\oL}} + \beta \;  {\exp(-i\int_b^{\bt} \oL \d
{\bt})\over \sqrt{\oL}}.
\end{equation}
Again, we will utilize the junction conditions to do so. 
From continuity of the scalar field, $[\phi]=0$, we obtain
\begin{equation}
   {1+i\over 2} \; {e^\Theta}
+   {1-i\over 2}\; {e^{-\Theta} } = {\alpha + \beta}.
\end{equation}
Continuity of the conjugate momentum, $[\dot{\phi}]=0$, gives us
\begin{equation}
 {1+i\over 2} \; e^\Theta -   {1-i\over 2} \; e^{-\Theta}  =
i(\alpha-\beta).
\end{equation}

\subsection{Bogoliubov coefficients:}
Solving these two simultaneous linear equations for $\alpha$ and
$\beta$ we find
\begin{equation}
\alpha = \cosh\Theta; \qquad \beta = i \sinh \Theta.
\end{equation}
Earlier we defined $\Theta = +\int_a^b \oE \, \d {\bt}$, and so
\begin{equation}
\alpha = \cosh\left\{+\int_a^b \oE \,\d {\bt}\right\};
\qquad
\beta = i \sinh \left\{+\int_a^b \oE \,\d {\bt} \right\}\,.
\end{equation}
Recalling that $\oE^{2}=A(\bar t)^{d} \left(m^2+k^2/A(\bar
t)\right)$, we find that the cosmological particle production for
an expanding or contracting spatially flat FLRW universe with a
region of Euclidean signature is governed by the Bogoliubov coefficient
\begin{equation}
\beta= i \sinh\left[ \int_E A(\bar
t)^{d/2} \sqrt{m^2+k^2/A(\bar t)} \;\d \bar t\right],
\end{equation}
which we can also rewrite as
\begin{equation}
\beta= i \sinh\left[ \int_E   \sqrt{m^2+k^2/A(\bar t)} \;\sqrt{|\bar B(\bar t)|}\;\d \bar t\right].
\end{equation}
In terms of our original $t$ coordinate, we can write this as
\begin{equation}
\beta= i \sinh\left[ \int_E \; \sqrt{m^2+k^2/A(t)} \;  \sqrt{|B(t)|}   \; \d t\right].
\end{equation}
Now introducing the physical scale factor via $A(t)\to a(t)^2$, and proper (Euclidean) time via $\d \tau =  \sqrt{|B(t)|}   \; \d t = \sqrt{|\bar B(\bar t)|}\;\d \bar t$, we have
\begin{equation}
\beta= i \sinh\left[ \int_E \; \sqrt{m^2+k^2/a(\tau)^2} \;   \; \d \tau\right].
\end{equation}
Finally, inserting appropriate factors of $c_0$,
\begin{equation}
\beta= i \sinh\left[ \int_E \; \sqrt{m^2 c_0^4+k^2 c_0^2/a(\tau)^2} \;   \; \d \tau\right],
\end{equation}
which is our promised result. Similarly
\begin{equation}
\alpha= \cosh\left[ \int_E \; \sqrt{m^2 c_0^4+k^2 c_0^2/a(\tau)^2} \;   \; \d \tau\right].
\end{equation}

\section{Consistency check: $E={1\over2} V$ for a rectangular barrier}
As a simple check on the calculation, it is a classic and quite standard result that for quantum tunnelling through a rectangular barrier one has (see for example~\cite[p.\ 79]{Landau}, or~\cite{Merzbacher, bounds})
\begin{equation}
T= { 4E (V-E)\over 4E(V-E) + V^2 \sinh^2[ \sqrt{2m(V-E)}\; L]},
\end{equation}
where $V$ is the height of the barrier, $L$ is its width, and $E$ is the incident energy.
If we now take the special case $E={1\over2} V$ we have
\begin{equation}
T={1\over 1+\sinh^2[ \sqrt{2m(V-E)}\; L]}= {1\over\cosh^2[ \sqrt{2m(V-E)}\; L]},
\end{equation}
from which, using the standard equivalencies
\begin{equation}
|\alpha|^2 \leftrightarrow 1/T
\qquad\hbox{  and } \qquad
 |\beta|^2 \leftrightarrow (1-T)/T = R/T,
\end{equation}
we see
\begin{equation}
|\alpha|^2 \leftrightarrow \cosh^2[ \sqrt{2m(V-E)}\; L] = \cosh^2[\kappa \; L]  \leftrightarrow 
 \cosh^2\left\{\;\left |\bar \omega\right|  \; \Delta \tau\right\}.
\end{equation}
That is
\begin{equation}
|\alpha| \leftrightarrow
 \cosh\left\{\;\left |\bar \omega\right|  \; \Delta \tau\right\},
 \qquad\qquad
|\beta| \leftrightarrow
 \sinh\left\{\;\left |\bar \omega\right|  \; \Delta \tau\right\},
\end{equation}
completely in agreement with our exact and WKB calculations.

%
\section{Quantum gravity (phenomenology) and signature change events?\label{Sec:QGP}}
%
There are several reasons for wishing to look at issues related to ``quantum gravity phenomenology", reasons such as possible Lorentz symmetry breaking, the limits of validity of the semi-classical gravity approximation, and the question of back-reaction.
Regarding Lorentz symmetry breaking --- one particularly important reason for looking at this topic is that it is known, in BEC-based analogue spacetimes, that the same physics that is responsible for Lorentz symmetry breaking is \emph{also} responsible for regularising particle production, and in fact rendering the number of particles produced finite. It would be interesting to see if Lorentz symmetry breaking regularises the particle production in this more abstract context. (In fact it does not, and in the present context we shall soon see that regularisation of the particle production requires more than just generic Lorentz symmetry breaking, it requires a particular form of Lorentz symmetry breaking.)
Beyond this,  the semi-classical gravity approximation, and the question of back-reaction, are central to the question of whether it is possible to define a \emph{dynamics} for signature change events --- at the present level of analysis we have treated signature change kinematically, not yet making any direct  attempt at assigning a suitable dynamical theory to this process. While dynamical theories of signature change (or signature selection) have been considered in the past (\eg, see~\cite{Jeff-Greensite, Elizalde, Odinstov, Roberts}), the (infinite) particle production we have encountered in this article very strongly suggests that any realistic dynamical theory must be very strongly dependent on back-reaction, and that failure to include the possible back-reaction due to the specific processes discussed in this article will render any possible dynamical system moot.
\subsection{($d+1$) massive minimally coupled scalar field with Lorentz breaking}
There are numerous reasons, typically based on quantum gravity phenomenology, for thinking that it might be interesting to consider the possibility of an ultra-high-energy breakdown of Lorentz invariance~\cite{Mattingly, causal-sets, broken-BEC}.
Indeed, let us adopt a Lorentz-breaking formalism similar to that of Jacobson~\cite{Jacobson}. (See also~\cite{bill}.) Consider  a PDE of the form
\begin{equation}
\Delta_{d+1} \phi - F(-\Delta_d) \phi = m^2 \phi,
\end{equation}
where $\Delta_{d+1}$ is the spacetime D'Alembertian while $\Delta_d$ is a purely spatial D'Alembertian. We shall again adopt definitions  (\ref{E:def1})--(\ref{E:def2}).
The relevant D'Alembertian operators are now specified by
\begin{equation}
\Delta_{d+1} \,\phi = {1\over\sqrt{-g_{d+1}}} \partial_a \left( \sqrt{-g_{d+1}} \;g^{ab} \;\partial_b \phi \right);
\end{equation}
\begin{equation}
\Delta_{d} \,\phi = {1\over\sqrt{g_d}} \partial_i\left( \sqrt{g_d} \;g^{ij} \;\partial_j \phi \right).
\end{equation}
Then the relevant PDE reduces to the explicit form
\begin{eqnarray}
&&\partial_a \left(
\mathrm{diag}\left[ - \sqrt{A^d/B}, \sqrt{B A^{d-2}},  \sqrt{B A^{d-2}}, \dots \right]^{ab}\;\partial_b \phi
\right)
\\
&&
\qquad\qquad\qquad
-\sqrt{ B A^d} \; F(-A^{-1} \nabla^2) \phi  = \sqrt{ B A^d} \;m^2 \, \phi,
\nonumber
\end{eqnarray}
which we rewrite as
\begin{equation}
\fl - \partial_t \left( \sqrt{A^d/B} \; \partial_t \phi\right) +
\nabla \left( \sqrt{B A^{d-2}} \; \nabla\phi \right) - \sqrt{ B
A^d}F(-A^{-1} \nabla^2) \phi =  \sqrt{ B A^d} \; m^2 \, \phi,
\end{equation}
implying
\begin{equation}
\fl \partial_t \left( \sqrt{A^d/B} \; \partial_t \phi\right) = -
\left[ \sqrt{ B A^d} \; \left\{m^2 + F( A^{-1} k^2)\right\}+
\sqrt{B A^{d-2}} \; k^2  \right] \phi.
\end{equation}
Now use exactly the same transformation, reparameterising $t\to \bar t$, (and so setting $B=\epsilon\, A^d$) to make the LHS simple.
One can just transcribe the previous discussion with minor modifications:
\begin{equation}
\partial_{\bar t}^2 \phi = - \epsilon \; A^d \left[ m^2 + F(k^2/A) + k^2/A \right] \phi.
\end{equation}
That is, we now have
\begin{equation}
\bar\omega^2 = \epsilon A^d \left[ m^2 + F(k^2/A) + k^2/A \right],
\end{equation}
where even in the Lorentzian signature region the presence of the $F(k^2/A)$ term indicates the presence of an explicit breaking of Lorentz invariance. Despite the presence of this Lorentz breaking term, the formulae for the Bogoliubov coefficients are only slightly altered.
Indeed, approximate WKB solutions for the mode functions in the Euclidean regime are now
\begin{equation}
\phi \approx {
\exp\left[\pm \displaystyle{\int} \sqrt{ m^2 + k^2/A + F(k^2/A)} \; A^{d/2} \d \bar t\right]
\over
\sqrt{  \sqrt{ m^2 + k^2/A + F(k^2/A) } \; A^{d/2} }
},
\end{equation}
leading, in a manner completely analogous to the preceding calculation, to the WKB estimates (up to irrelevant phases) for the Bogoliubov coefficients:
\begin{equation}
\alpha \approx\cosh\left\{ \int_E \sqrt{ m^2 + k^2/A + F(k^2/A) } \;A^{d/2} \; \d \bar t\right\};
\end{equation}
\begin{equation}
\beta \approx i \sinh\left\{ \int_E \sqrt{ m^2 + k^2/A  + F(k^2/A)} \;A^{d/2} \; \d \bar t\right\}.
\end{equation}
It is easy to check that this reproduces all the other cases we have considered in appropriate limits.
Explicitly introducing the standard scale factor $A(\bar t)\to a(\bar t)^2$, utilizing the dispersion relation in the form
\begin{equation}
\omega= \Omega(k) = \sqrt{ m^2 + k^2 + F(k^2)},
\end{equation}
and working in terms of (Euclidean) proper time $\tau$, we finally obtain the promised result:
\begin{eqnarray}
\alpha &\approx& \cosh\left\{ \lint_{\!\!\!\!\!\!E}  \; \Omega\!\left({k /a(\tau)}\right) \;\d \tau  \right\};
\\
\beta &\approx&  i\sinh\left\{  \lint_{\!\!\!\!\!\!E}  \; \Omega\!\left({k / a(\tau)}\right) \;\d \tau \right\}.
\end{eqnarray}
Note that the adoption of \emph{this particular type} of Lorentz symmetry breaking does not ameliorate the production of infinite numbers of particles, and infinite energy in the QFT, that we have encountered in previous calculations:
\begin{eqnarray}
N_\# &\approx& \oint 
\sinh^2\left\{ \lint_{\!\!\!\!\!\!E}  \; \Omega\!\left({k / a(\tau)}\right) \;\d \tau  \right\} \; \d^d k;
\\
E &\approx&  \oint \omega\!\left({k / a_\infty}\right) \; 
\sinh^2\left\{  \lint_{\!\!\!\!\!\!E}  \; \Omega\!\left({k / a(\tau)}\right) \;\d \tau \right\} \; \d^d k.
\end{eqnarray}
These infinities can be tracked back to the fact that in the current class of models we have retained full coordinate invariance for reparameterisations of the time variable, so that on transforming $t\to\bar t$ the value for $\bar B$ was specifically chosen to make it equal to $A^d$, and so effectively eliminate it from the PDE for the mode functions. By invoking the time reparameterisation invariance present in general relativity, and then suitably choosing the time coordinate, we have greatly simplified the calculation (rendering it tractable) at the cost of limiting the class of Lorentz violations one can consider.

This is emphatically \emph{not} what happens in analogue spacetimes, where the presence of a background laboratory time explicitly breaks time reparameterisation invariance and greatly influences the  physics. Consider, for example, the Bogoliubov coefficients as presented in~\cite{silke+angela}. There the authors present a real-life model, one that is capable of undergoing signature changes in the geometry emergent from an ultra-cold gas of weakly interacting Bosons, and where the Bogoliubov coefficients are of the form
\begin{equation}
\alpha \approx\cosh\left\{ \int_E \sqrt{ -B\,m^2 - \varepsilon_{\mathrm{qp}}^2 \, k^4 -B\,  k^2/A } \;A^{d/2} \; \d \bar t\right\};
\end{equation}
\begin{equation}
\beta \approx i \sinh\left\{ \int_E \sqrt{ -B\,m^2 - \varepsilon_{\mathrm{qp}}^2 \, k^4 - B\, k^2/A } \;A^{d/2} \; \d \bar t\right\},
\end{equation}
such that the extreme ultraviolet modes are \emph{unaffected} by the sign change of $B$. (The constant $\varepsilon_{\mathrm{qp}}=\hbar/(2M)$ in front of the $k^4$ term is in this situation determined by the physical mass $M$ of the bosons that undergo condensation, and is emphatically not affected by signature change.) Sufficiently high energy modes will maintain their oscillatory behaviour, beyond the region of Riemannian signature.
In some sense the introduction of Lorentz symmetry breaking in the current formalism has made the situation worse --- to regularise the number and total energy of the particles produced in the signature change it seems that one must do something much more drastic than just break Lorentz symmetry: It seems that one needs to break time reparameterisation invariance, since only then can one hope to give an explicit  and invariant meaning to the idea that the ``amount'' of signature change, quantified by $B(t)$, might be ``small''. As long as one retains time reparameterisation invariance then the magnitude of $B(t)$, though not its sign, is coordinate dependent; and asking that signature change be ``small'' is a meaningless request.

\subsection{Dynamics of signature change events\label{Sec:Dynamics}}
Regarding the possible dynamics of signature change events, it is important to realise that in standard general relativity the signature is not a local dynamical variable --- at best it is a ``non-local'' and discrete dynamical variable associated with entire regions,  rather than a dynamical field associated with specific points on the manifold. Specifically, in the standard Einstein--Hilbert implementation of general relativity the lapse and shift show up as Lagrange multipliers enforcing the super-Hamiltonian and super-momentum constraints, and so are explicitly non-dynamical. This observation may however be altered in certain modified theories of gravity,  specifically in theories that break time reparameterisation invariance. Models such as BEC-based analogue spacetimes, because they are bi-metric, (they contain both a background spacetime metric and an ``acoustic metric'') explicitly break time reparameterisation invariance --- the ratio between the speed of sound and the speed of light is a dimensionless observable that can be given an explicit coordinate independent meaning and this dimensionless ratio certainly has dynamics (dependent on the details of the BEC system under consideration). Effectively the lapse function of the acoustic metric has acquired a dynamics due to the presence of the background (laboratory) spacetime.

If (and this is by no means certain) the true physics of ``quantum gravity'' is ultimately based on a bi-metric theory, then similarly the low-energy lapse function associated with ``our'' spacetime metric is likely to acquire dynamics, but only in those situations where the coupling of  ``our'' spacetime metric to the ``other'' (background) spacetime metric becomes non-negligible.  However --- and this is important --- given that our naive calculations (which ignored back-reaction) have resulted in infinite particle production and infinite energy release, it becomes clear that (barring miracles) any attempt at including back-reaction should lead to a strong coupling between the two metrics of any bi-metric theory --- at which stage the signature becomes dynamical, one leaves the semi-classical regime, and the theory of ``quantum gravity'' must be invoked.

It is in this sense that signature change is ultimately a challenge for the as yet unknown theory of quantum gravity itself --- attempting to analyse it within the framework of curved-spacetime QFT, or even semi-classical gravity, has led to an impasse. Like the questions related to topology change~\cite{topology}, and ``chronology protection'' (the avoidance of closed timelike curves)~\cite{chronology}, it seems that detailed analysis of this issue inexorably moves one outside the framework of semi-classical gravity --- it is ultimately  ``quantum gravity'' (whatever that may be) that one will have to face in order to definitively answer the questions we raise in this article.

%
\section{Discussion and Conclusions\label{Sec:Discussion}}
%
The possibility of signature change in ordinary general relativity is a topic that in the past has lead to considerable interest~\cite{no-boundary, Hartle-Hawking, arrow, Dray-et-al, Dray-other, Kandrup} and often controversy~\cite{controversy1, controversy2}.  In this article we have made a contribution to the theory of cosmological particle production due to a Lorentzian$\to$Euclidean$\to$Lorentzian signature change.
\begin{itemize}
\item One of the key technical points of the article is the realization that cosmological particle production in signature change is closely related to quantum tunnelling ``half-way up the barrier''.
\item The WKB estimates we have derived can very easily be generalized to provide estimates for quantum tunneling through sharp walled barriers (with generic complicated peaks and valleys, as long as the allowed $\leftrightarrow$ forbidden transition occurs at the sharp vertical wall).
\item
We have seen that generically signature change seems to drive the production of an infinite number of particles, with infinite total energy, and that these infinities are not ameliorated by dimension, rest mass, or even the most physically reasonable sub-class of Lorentz symmetry violations. 

\item 
If one wishes to retain signature change as a physically viable process then it really looks like one should be searching for a suitable dynamics to both drive the signature change event \emph{and} to regularise both the  \emph{number} and \emph{energy} of the resulting particle emission. The only currently known mechanism for doing so is based on the presence of a preferred background time, of the type occurring in the so-called ``analogue spacetimes''.  Of course the implied breakdown of time reparameterisation invariance will be deeply disturbing to classical general relativists.

\item One way of phrasing our results is this: Quantum field theory reacts violently to any attempt at imposing signature change on the spacetime manifold. In a manner reminiscent of the way that quantum field theory reacts violently to any attempt at imposing topology change~\cite{topology}, or chronology violations~\cite{chronology}, imposing an externally driven signature change seems to catapult  one into the realm of ``quantum gravity''.

\end{itemize}
Overall, we have seen that computational and conceptual advances in signature changing spacetimes can still be made, with what is from a technical perspective a remarkably straightforward and ``fundamental'' calculation.
\ack
AW was partially supported by the Australian Research Council and by  the Australian National University.
MV was supported by the Marsden Fund administered by the Royal Society of New Zealand.
SW was supported by an EU Marie Curie fellowship.

\section*{References}
\addcontentsline{toc}{section}{References}


\end{document}